\documentclass[amssymb,prb,twocolumn,showpacs]{revtex4}
\input{psfig.sty}
\usepackage{epsfig}
\usepackage{dcolumn}
\usepackage{amsmath}
\begin{document}
\title{\bf\ Theoretical approach to microwave radiation-induced
zero-resistance states in 2D electron systems}
\author{J. I\~narrea$^{1,2}$
 and G. Platero$^1$}
\affiliation{$^1$Instituto de Ciencia de Materiales,
CSIC, Cantoblanco, Madrid, 28049, Spain \\
$^2$Escuela Polit\'ecnica
Superior, Universidad Carlos III, Leganes, Madrid.}
\date{\today}
\begin{abstract}
We present a theoretical model in which the existence of
radiation-induced zero-resistance states is analyzed. An exact
solution for the harmonic oscillator wave function in the presence
of radiation, and a perturbation treatment for elastic scattering due
to randomly distributed charged impurities, form the foundations of
our model. Following this model most experimental results are
reproduced, including the formation of resistivity oscillations,
their dependence on the intensity and frequency of the
radiation, temperature effects, and the locations of the resistivity
minima. The
existence of zero-resistance states is thus explained in terms of the
interplay of the electron MW-driven orbit dynamics and the Pauli
exclusion principle.
\end{abstract}
\maketitle
 In the last two decades, especially since the discovery of
the Quantum Hall Effect, a lot of progress has been made in
the study of two-dimensional electron systems (2DES), and very
important and unusual properties have been discovered when these
systems are subjected to external AC and DC fields.
In the last two years two experimental
groups\cite{mani,zudov} have announced the existence of vanishing
resistance in 2DES, i.e. zero resistance states (ZRS), when these
systems are under the influence of a moderate magnetic field
($B$) and microwave (MW) radiation simultaneously. In the same kind
of experiments large resistivity oscillations have
been observed\cite{mani,zudov,potemski1,potemski2}.\\
 The discovery of this novel effect has led to a great deal of
theoretical activity, and  among the most interesting contributions we
can summarize various different approaches. Some, like Girvin et al,
\cite{girvin,lei} argue that this striking effect has to do with
photon-assisted scattering from impurities or disorder, or
alternatively arises from acoustic phonon scattering \cite{ryz2}.
Others \cite{shi,rivera} relate the ZRS with a new structure of
the density of states of the system in the presence of light.
According to Andreev's approach\cite{andreev}, the key is the
existence of an inhomogeneous current flowing through the sample
due to the presence of a domain structure in it. To date there is
no consensus about the true origin. All the theories above have in
common that they predict negative resistivity, while this has not
been experimentally confirmed.
 Only Willett et al \cite{willett}, have recently
observed negative conductivity for certain
configurations of contacts.\\
 In this Letter  we
develop a semi-classical model that is based on the exact solution
of the electronic wave function in the presence of a static $B$
interacting with MW-radiation, i.e. a quantum forced harmonic
oscillator, and a perturbation treatment for elastic scattering
from randomly distributed charged impurities. We explain and
reproduce most experimental features, and clarify the physical
origin of ZRS. For large MW field amplitudes, final states for
electrons, which semi-classically describe orbits whose center
positions oscillate due to the MW field, will be occupied. The MW
field thus blocks the electron movement between orbits, and the
longitudinal conductivity and resistivity $\rho_{xx}$ will be zero
(see Fig.1). We explain the dependence of $\rho_{xx}$ on
temperature ($T$) by means of the electron interaction with
acoustic phonons which acts as a damping factor for the forced
quantum oscillators, and we discuss the ZRS dependence on $B$. We
observe them at $w/w_{c}=j+1/4$ for the experimental parameters of
ref.[1], being $w$ the MW frequency, $w_{c}$ the cyclotron
frequency and $j$ an integer. In our model we do not consider the
induced electrostatic potentials by the charge distribution within
the sample, nor the dynamical electronic response induced by the
AC-potential.\cite{gloria,butti}\\
 We first obtain
an exact expression of the electronic wave vector for a 2DES in a
perpendicular $B$, a DC electric field and MW radiation which is
considered semi-classically.
 The total hamiltonian $H$ can be written as:
\begin{eqnarray}
H&=&\frac{P_{x}^{2}}{2m^{*}}+\frac{1}{2}m^{*}w_{c}^{2}(x-X)^{2}-eE_{dc}X +\nonumber \\
 & &+\frac{1}{2}m^{*}\frac{E_{dc}^{2}}{B^{2}}\nonumber-eE_{0}\cos wt (x-X) -\nonumber \\
 & &-eE_{0}\cos wt X \nonumber\\
 &=&H_{1}-eE_{0}\cos wt X
\end{eqnarray}
 $X$ is the center of the orbit for the electron spiral motion:
$X=\frac{\hbar k_{y}}{eB}- \frac{eE_{dc}}{m^{*}w_{c}^{2}}$,
$E_{0}$ the intensity for the MW field and $E_{dc}$ is the DC
electric field in the $x$ direction. $H_{1}$ is the hamiltonian
corresponding to a forced harmonic oscillator whose orbit is
centered at $X$. $H_{1}$ can be solved exactly \cite{kerner,park},
and using this result allows an exact solution for the electronic
wave function of $H$ to be obtained:
\begin{eqnarray}
&&\Psi(x,t)=\phi_{n}(x-X-x_{cl}(t),t)\nonumber  \\
&&\times  exp \left[i\frac{m^{*}}{\hbar}\frac{dx_{cl}(t)}{dt}[x-x_{cl}(t)]+
\frac{i}{\hbar}\int_{0}^{t} {\it L} dt'\right]\nonumber  \\
&&\times\sum_{m=-\infty}^{\infty} J_{m}\left[\frac{eE_{0}}{\hbar}
X\left(\frac{1}{w}+\frac{w}{\sqrt{(w_{c}^{2}-w^{2})^{2}+\gamma^{4}}}\right)\right]
e^{imwt}
\end{eqnarray}
where $\gamma$ is a phenomenologically-introduced damping factor
for the electronic interaction with acoustic phonons, $\phi_{n}$
is the solution for the Schr\"{o}dinger equation of the unforced
quantum harmonic oscillator and $x_{cl}(t)$ is the classical
solution of a forced harmonic oscillator\cite{park},
$x_{cl}=\frac{e E_{o}}{m^{*}\sqrt{(w_{c}^{2}-w^{2})^{2}+\gamma^{4}}}\cos wt$.
${L}$ is the classical lagrangian, and $J_{m}$ are Bessel
functions. Apart from phase factors, the wave function for $H$ is
the same as the standard harmonic oscillator where the center is
displaced by $x_{cl}(t)$. Now we introduce the impurity scattering
suffered by the electrons in our model \cite{ridley}. If the
scattering is weak we can apply time dependent first order
perturbation theory, starting from $H$ as an exact hamiltonian and
$\Psi_{l}(x,t)$ as the wave-vector basis. The aim is to calculate
the transition rate from an initial state $\Psi_{n}(x,t)$, to a
final state $\Psi_{m}(x,t)$:
\begin{equation}
W_{n,m}=\lim_{\alpha\rightarrow 0} \frac{d}{d t} \left|
 \frac{1}{i \hbar} \int_{-\infty}^{t^{'}}<\Psi_{m}(x,t) |V_{s}|\Psi_{n}(x,t)>e^{\alpha t}d t\right|^{2}
\end{equation}
where $V_{s}$ is the scattering potential for charged impurities\cite{ando}:
$V_{s}= \sum_{q}\frac{e^{2}}{2 S \epsilon (q+q_{s})} \cdot e^{i
\overrightarrow{q}\cdot\overrightarrow{r}}$,
$S$ being the surface of the sample, $\epsilon$ the dielectric
constant and $q_{s}$ is the Thomas-Fermi screening
constant\cite{ando}.
After some lengthy algebra we
 arrive at the following expression for the transition rate:
\begin{eqnarray}
&&W_{n,m}= \frac{e^{5}n_{i}B S}
{16\pi^{2}\hbar^{2}\epsilon^{2}}\left[\frac{\Gamma}{[\hbar w_{c}(n-m)]^{2}+\Gamma^{2}}\right]\nonumber\\
 &&\times\int_{0}^{q_{max}}dq
\frac{q}{(q^{2}+q_{0}^{2})^{2}} \frac{n_{1}!}{n_{2}!}e^{-\frac{1}{2}q^{2}R^{2}}\left(\frac{1}{2}q^{2}R^{2}\right)^{n_{1}-n_{2}}\nonumber \\
&&\times \left[L_{n_{2}}^{n_{1}-n_{2}}(
\frac{1}{2}q^{2}R^{2})\right]^{2}J_{0}^{2}(A_{m})J_{0}^{2}(A_{n})
\end{eqnarray}
where $A_{n(m)}= \frac{eE_{0}}{\hbar}
X_{n(m)}\left(\frac{1}{w}+\frac{w}{\sqrt{(w_{c}^{2}-w^{2})^{2}+\gamma^{4}}}\right)$.
With the experimental parameters we take, the arguments of
the Bessel functions are very small $(\sim 10^{-2})$, and only
$J_{0}$ terms need to be considered. $\Gamma$ is the Landau level
broadening, $n_{i}$ is the impurity density and $R$ is the magnetic
characteristic length $R^{2}=\frac{\hbar}{eB}$.
$L_{n_{2}}^{n_{1}-n_{2}}$ are the associated Laguerre polynomials,
$n_{1}= max (n,m)$ and $n_{2}= min (n,m)$.\\
Without radiation, an electron in an initial state $\Psi_{n}$
corresponding to an orbit center position $X_{n}^{0}$, scatters and
jumps to a final state $\Psi_{m}$ with orbit center $X_{m}^{0}$,
changing its average coordinate in the static electric field
direction by $\Delta X^{0}=X_{m}^{0}-X_{n}^{0}=q \cos \theta R^{2}$
(polar coordinates, $q$ and $\theta$, have been used). In the
presence of MW radiation, the electronic orbit center coordinates change and
are given according to our model by $X^{MW}=X^{0}+x_{cl}(t)$.
This means that due to the MW field all the electronic orbit
centers in the sample oscillate back and forth in the $x$ direction
through $x_{cl}$.
\begin{figure}
\centering\epsfxsize=3.5in \epsfysize=3.5in
\epsffile{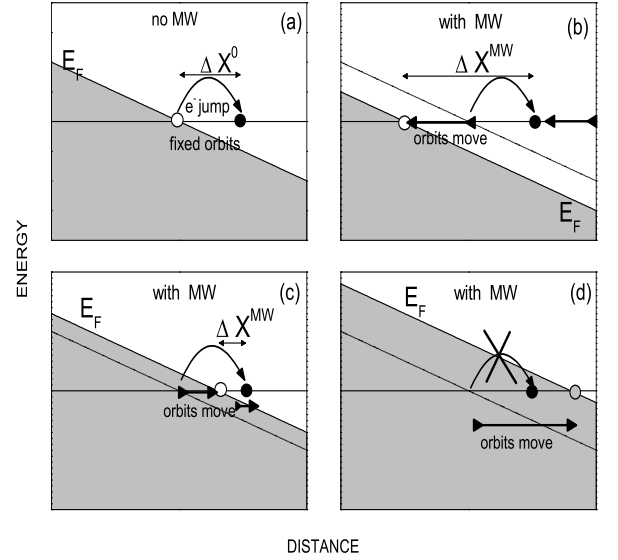}
\caption{Schematic diagrams of electronic transport without and
with MW. In Fig.(1.a) no MW field is present, and due to
scattering electrons jump between fixed-position orbits. When the
MW field is on, the orbits are not fixed but oscillate at $w$. In
Fig.(1.b) the orbits move backwards during the jump, and on
average electrons advance further than in the no MW case. In
Fig.(1.c) the orbits are moving forwards, and on average electrons
advance less than in the no MW case. In Fig. (1.d) the orbits are
moving forwards but their amplitudes are larger than the
electronic jump, and the electron movement between orbits cannot
take place because the final state is occupied. This situation
corresponds to ZRS.}
\end{figure}
We have to consider two important factors: first, when an electron
suffers a scattering process with a probability given by $W_{m,n}$,
it takes a time $\tau = \frac{1}{W_{m,n}}$ for that electron to jump
from an orbit center to another. Secondly, in the jump, as in any
other scattering event, the electron loses memory and phase with
reference to the previous situation, and thus when it arrives at the final
state the oscillation condition is going to be different from the
starting point. If we consider that the
oscillation is at its {\em mid-point} when the
electron jumps from the initial state, and that it takes a time
$\tau$ to get to the final one, then we can write for the average
coordinate change in the $x$ direction: $\Delta X^{MW}=\left(\Delta
X^{0}+ \frac{e
E_{o}}{m^{*}\sqrt{(w_{c}^{2}-w^{2})^{2}+\gamma^{4}}}\cos
w\tau\right)$.\\
In Fig. (1) we present schematic diagrams for the different
situations. In Fig. (1.a) no MW field is present and electrons
jump between fixed orbits, and on average an electron advances a
distance $\Delta X^{0}$. When MW field is on, the orbits are not
fixed, and instead move back and forth through $x_{cl}$. Three
cases can be distinguished. In Fig. (1.b) the orbits are moving
backwards during the electron jump, and on average, due to
scattering processes, the electron advances a larger distance than
in the no MW case, $\Delta X^{MW}>\Delta X^{0}$. This corresponds
to an increasing conductivity. In Fig. (1.c) the orbits are moving
forwards and the electron advances a shorter distance, $\Delta
X^{MW}<\Delta X^{0}$. This corresponds to a decrease in the
conductivity with respect to the case without MW. If we increase
the MW intensity, we will eventually reach the situation depicted
in Fig. (1.d) where orbits are moving forwards, but their
amplitude is larger than the electronic jump. In that case the
electronic jump is blocked by the Pauli exclusion principle
because the final state is occupied. This is the physical origin
of ZRS. At different ranges of parameters, i.e. larger MW power or
smaller $w$, additional terms corresponding to Bessel functions of
order higher than zero would contribute to Eq. 4,
which may eventually produce negative conductivity\cite{willett,ina}.\\
If the average value $\triangle X^{MW}$ is different from zero
over all the scattering processes, the electron possesses an
average drift velocity $v_{n,m}$ in the $x$
direction\cite{ridley}. This drift velocity can be calculated
readily by introducing the term $\triangle X^{MW}$ into the
integrand of the transition rate, and finally the longitudinal
conductivity $\sigma_{xx}$ can be written as: $\sigma_{xx}=
\frac{2e}{E_{dc}}\int \rho(E_{n}) v_{n,m}
[f(E_{n})-f(E_{m})]dE_{n}$. Gathering all the terms, we finally
obtain the expression:
\begin{eqnarray}
 &&\sigma_{xx}(E_{n})=\frac{e^{7}n_{i}B^{2}S}{16\pi^{5}\epsilon^{2}\hbar
 ^{3}E_{dc}}\sum_{n,m}\left[\frac{\Gamma}{[\hbar w_{c}(n-m)]^{2}+\Gamma^{2}}\right]\nonumber\\
&& \times \int dE_{n}\left[\frac{\Gamma}{[E_{n}-\hbar
w_{c}(n+\frac{1}{2})]^{2}+\Gamma^{2}}\right]
\left[f(E_{n})-f(E_{m})\right]\nonumber\\
&&\times\int_{0}^{q_{m}}dq \left[\frac{q( q \cos \theta R^{2}+A\cos w \tau)}{(q+q_{s})^{2}}\right]\nonumber \\
&&\times   \frac{n_{1}!}{n_{2}!}e^{-\frac{1}{2}q^{2}R^{2}}\left(
\frac{1}{2}q^{2}R^{2}\right)^{n_{1}-n_{2}}\left[L_{n_{2}}^{n_{1}-n_{2}}( \frac{1}{2}q^{2}R^{2})\right]^{2}\nonumber\\
&& \times J_{0}^{2}(A_{m})J_{0}^{2}(A_{n})
\end{eqnarray}
\begin{figure}
\centering \epsfxsize=3.5in \epsfysize=3.0in
\epsffile{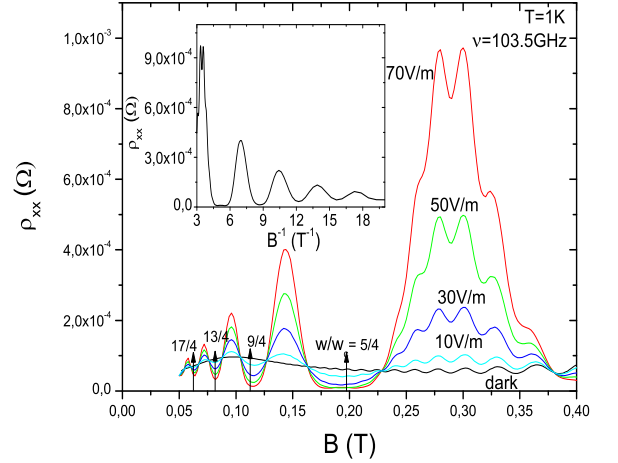}
\caption{Calculated magnetoresistivity $\rho_{xx}$ as a function
of $B$, for different MW intensities but for the same frequency
$\nu=103.5  GHz$. The darkness case is also presented. In the
inset we show $\rho_{xx}$ vs $B^{-1}$, which is roughly periodic
in $B^{-1}$ (T=1K).}
\end{figure}
\begin{figure}
\centering \epsfxsize=3.5in \epsfysize=2.3in
\epsffile{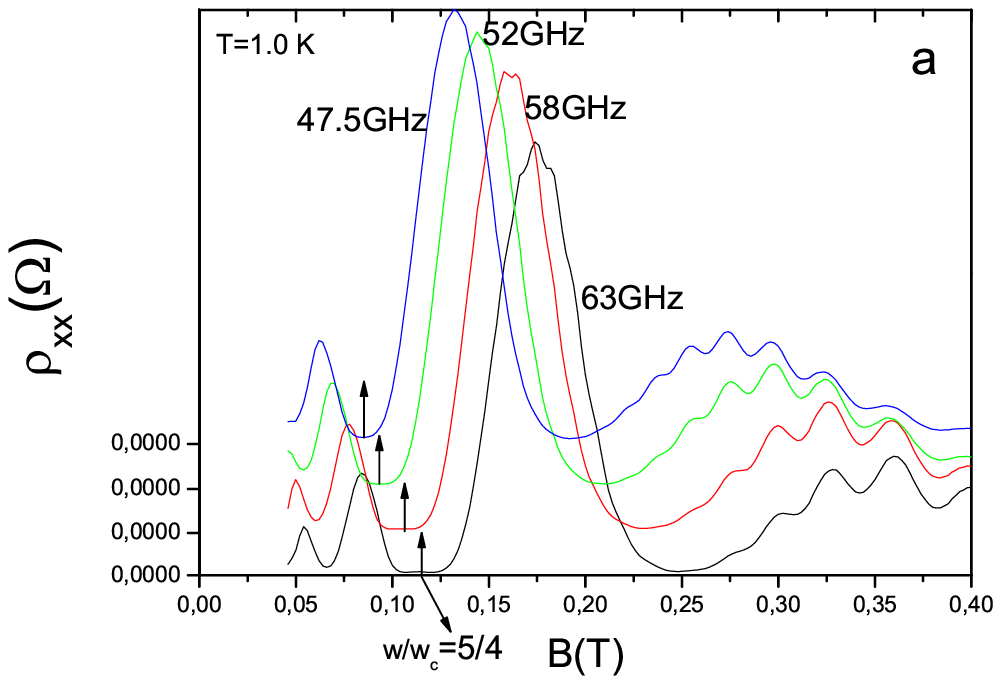} \centering \epsfxsize=3.5in
\epsfysize=2.3in \epsffile{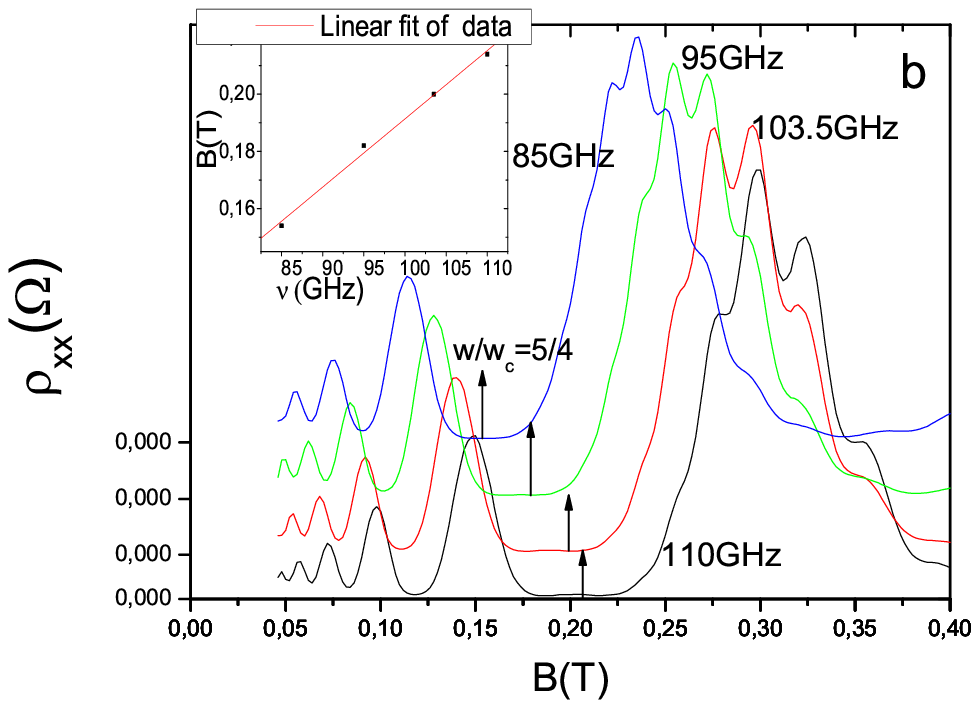}
\caption{Magnetoresistivity response for different MW frequencies. a)
corresponds to a range of low $w$ values and b) to higher ones. In
both cases, ZRS are reproduced (T=1K).}
\end{figure}
where the density of states $ \rho(E_{n})$ has been simulated by a
Lorentzian function, being $A=\frac{e
E_{o}}{m^{*}\sqrt{(w_{c}^{2}-w^{2})^{2}+\gamma^{4}}}$. To obtain
$\rho_{xx}$ we use the relation
$\rho_{xx}=\frac{\sigma_{xx}}{\sigma_{xx}^{2}+\sigma_{xy}^{2}}
\simeq\frac{\sigma_{xx}}{\sigma_{xy}^{2}}$, where
$\sigma_{xy}\simeq\frac{n_{i}e}{B}$ and
$\sigma_{xx}\ll\sigma_{xy}$. All our results have been based on
experimental parameters corresponding to the experiments of Mani
\cite{mani} et al. In Fig. 2 we show the magnetoresistivity
$\rho_{xx}$ obtained using our model, as a function of $B$ for
different MW field intensities, in all cases using the same
frequency $w/2\pi=\nu=103.5$ GHz. The darkness case is also
presented. As the field intensity is lowered, the $\rho_{xx}$
response decreases to eventually reach the darkness response. In
the inset it is possible to see the calculated $\rho_{xx}$ vs
$B^{-1}$, which is roughly periodic in $B^{-1}$ in agreement with
experiment. The minima positions as a function of $B$ are
indicated with arrows, corresponding to
$\frac{w}{w_{c}}=j+\frac{1}{4}$. In the minima corresponding to
$j=1$, ZRS are found. Although the qualitative behavior of
$\rho_{xx}$ as a function of $B$ is very similar to the
experimental one, the absolute value is smaller. It could be due
to the smaller carrier density, or the simplified model for the
electronic scattering with impurities that we have considered.
\begin{figure}
\centering \epsfxsize=3.5in \epsfysize=3.0in
\epsffile{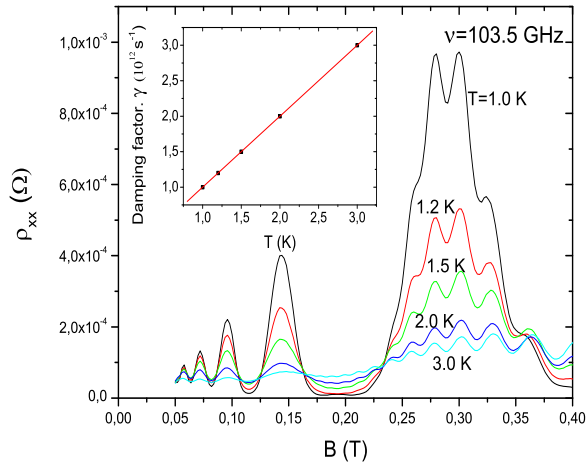}
\caption{ $\rho_{xx}$ versus B for different temperatures with constant power
excitation. The oscillations get smaller as $T$ is increased, but
the positions of the minima stay constant. In the inset we show a
calculated linear response between $\gamma$ and $T$.}
\end{figure}
Fig. 3 shows $\rho_{xx}$ versus B for different MW frequencies.
The upper figure corresponds to a range of small values and the
lower figure to large ones. In both cases ZRS are reproduced very
clearly for $j=1$. Minima positions shift as $w$ is altered,
maintaining the ratio $\frac{w}{w_{c}}=j+\frac{1}{4}$. This is
shown in the inset, where the magnetic field B corresponding to
ZRS (j=1) for four different MW frequencies is plotted as a
function of $w$.
 As in experimental results presented in ref.
[1], a quite reasonable linear fit is achieved.\\
The dependence of $\rho_{xx}$ on $T$ (Fig. 4), has been obtained at
$\nu=103.5$ GHz. As $T$ increases $\rho_{xx}$ is softened, and
eventually almost disappears. The explanation can be readily
obtained through the damping parameter $\gamma$. When the
electronic orbits are oscillating harmonically due to the
time-dependent external force, interaction  with acoustic phonons
occurs. This interaction acts as a damping for the orbits'
movement. As $T$ increases, the lattice-orbit interaction
strengthens and the damping of orbit dynamics will be stronger as
well, giving a progressive reduction in the MW-induced $\rho_{xx}$
response. We have considered a linear dependence between $\gamma$
and $T$ as
in the experiments by Studenikin et al \cite{potemski1}.\\
Using our model, it is now possible to shed some light on the
peculiar dependence of $\rho_{xx}$ on $B$. According to our
calculations we have found an approximately linear relation between
$\rho_{xx}$ maxima and $B$, and we can therefore express the
corresponding dependence as: $\rho_{xx}\propto  B \cos(w
\tau)\propto B \cos\left(\frac{w}{B}\right)$.
Looking at the cosine argument ($w\tau$) it is clear that if we
change $w$, the minima positions will change as well. Regarding
$\tau=\frac{1}{W_{n,m}}$ we can say that the scattering transition
rate $W_{n,m}$ is mostly dependent on sample and scattering
variables, and that specific minima positions will be a function
of those variables. In this way we expect that for significantly
different samples, minima positions and other features of
$\rho_{xx}$ oscillations will change, explaining the discrepancy
observed between different experiments. \\
In summary, we have presented a new theoretical model whose main
foundations are the exact solution for the quantum harmonic
oscillator in the presence of MW radiation
and elastic scattering due to randomly distributed charged
impurities. This model gives a description of the electronic orbit
dynamics which is crucial to explain the physical origin of ZRS.
We are able to reproduce most experimental results, including
$\rho_{xx}$ oscillations, minima positions and their dependence on
$w$, MW intensity and $T$.\\
We acknowledge K. von Klitzing, J.H. Smet, T. Brandes, C. E.
Creffield, D. S\'anchez and R. L\'opez for enlightening
discussions and critical reading of the manuscript. This work was
supported by the MCYT (Spain) grant MAT2002-02465, the ``Ramon y
Cajal'' program (J.I.). and the EU Human Potential Programme
HPRN-CT-2000-00144.

\end{document}